\documentstyle[prl,aps,preprint,amssymb,epsfig,floats]{revtex}

\begin{document}

\def\d{{\rm d}}
\def\e{{\rm e}}
\def\O{{\rm O}}
\def\half{\mbox{$\frac12$}}
\def\eref#1{(\protect\ref{#1})}
\def\etal{{\it{}et~al.}}

\setcounter{topnumber}{2}
\renewcommand{\topfraction}{0.9}
\renewcommand{\textfraction}{0.1}
\renewcommand{\floatpagefraction}{0.5}

\newdimen\captwidth
\captwidth=5.5in
\def\capt#1{\refstepcounter{figure}\bigskip\hbox to \textwidth{%
       \hfil\vbox{\hsize=\captwidth\renewcommand{\baselinestretch}{1}\small
       {\sc Figure \thefigure}\quad#1}\hfil}\bigskip}

\title{Mean-field solution of the small-world network model}
\author{M. E. J. Newman, C. Moore, and D. J. Watts}
\address{Santa Fe Institute, 1399 Hyde Park Road, Santa Fe, NM 87501}
\maketitle

\begin{abstract}
  The small-world network model is a simple model of the structure of
  social networks, which simultaneously possesses characteristics of both
  regular lattices and random graphs.  The model consists of a
  one-dimensional lattice with a low density of shortcuts added between
  randomly selected pairs of points.  These shortcuts greatly reduce the
  typical path length between any two points on the lattice.  We present a
  mean-field solution for the average path length and for the distribution
  of path lengths in the model.  This solution is exact in the limit of
  large system size and either large or small number of shortcuts.
\end{abstract}

\newpage

Social networks, such as networks of friends, have two characteristics
which one might imagine were contradictory.  First, they show
``clustering,'' meaning that two of your friends are far more likely also
to be friends of one another than two people chosen from the population at
random.  Second, they exhibit what has become known as the ``small-world
effect,'' namely that any two people can establish contact by going through
only a short chain of intermediate acquaintances.  Following the work of
Milgram\cite{Milgram67}, it is widely touted that the average number of
such intermediates is about six---there are ``six degrees of separation''
between two randomly chosen people in the world.  In fact this number is
probably not a very accurate estimate, but the basic principle is sound.

\begin{figure}
\begin{center}
\psfig{figure=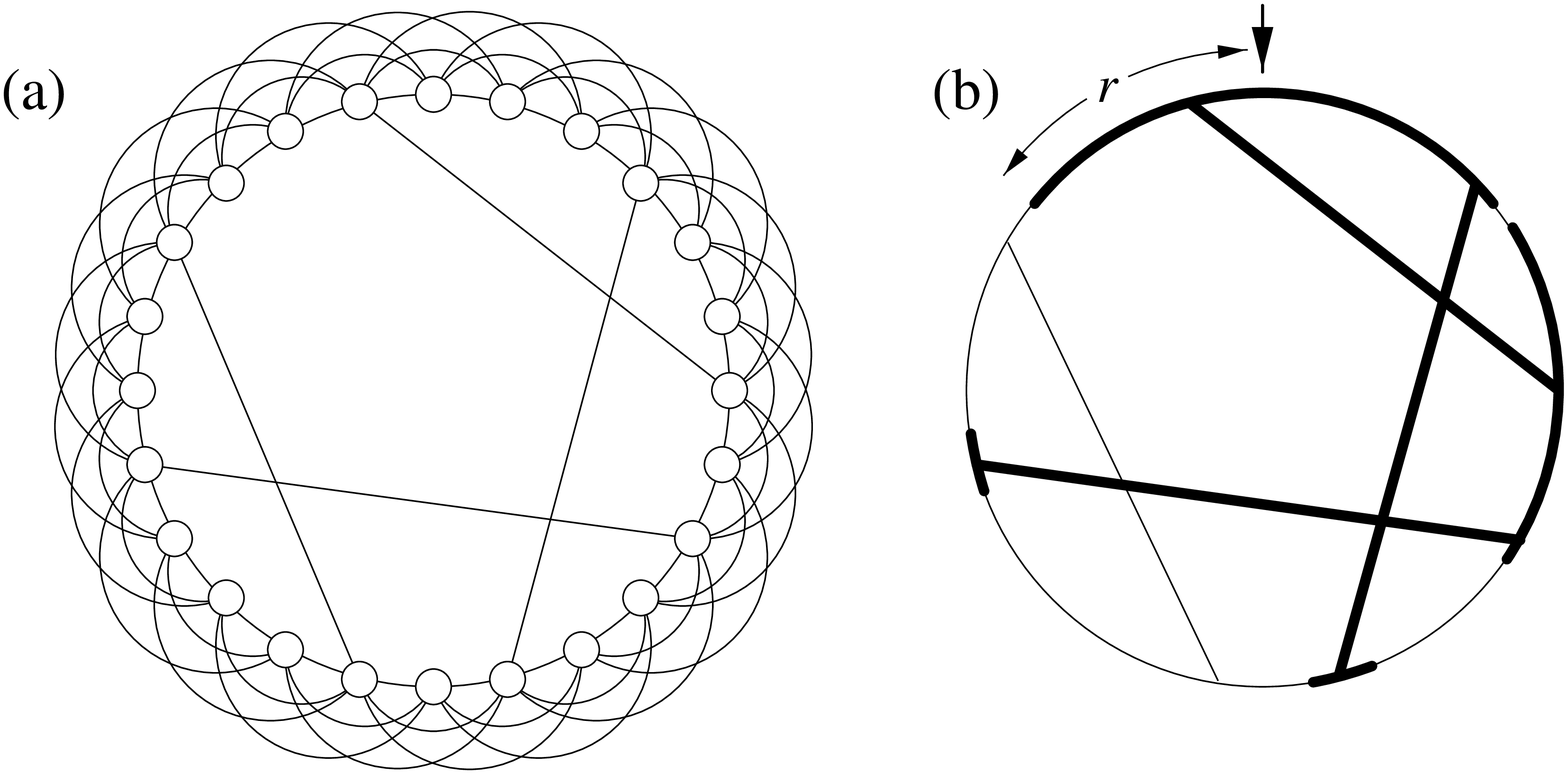,width=5in}
\end{center}
\capt{(a)~A small-world graph of 24 sites with $k=3$ and four shortcuts.
  (b)~The continuum version of the same graph.  The bold lines denote the
  portion of the graph which is within distance $r$ of the point at the top
  denoted by the arrow.  In this case there are four filled segments, or
  ``clusters'', around the perimeter of the graph, or equivalently four
  gaps between clusters. }
\label{model}
\end{figure}

These two properties appear contradictory because the first is a typical
property of low-dimensional lattices but not of random graphs or other
high-dimensional lattices, while the second is typical of random graphs,
but not of low-dimensional lattices.  Recently, Watts and
Strogatz\cite{WS98} have proposed a simple model of social networks which
interpolates between low-dimensional lattices and random graphs and
displays both the clustering and small-world properties.  In this model,
$L$ sites are placed on a regular one-dimensional lattice with nearest- and
next-nearest-neighbor connections out to some constant range $k$ and
periodic boundary conditions (the lattice is a ring).  A number of
shortcuts are then added between randomly chosen pairs of sites with
probability $\phi$ per connection on the underlying lattice (of which there
are $Lk$).  Thus there are on average $Lk\phi$ shortcuts in the graph.  An
example of a small-world graph with $L=24$, $k=3$, and four shortcuts is
shown in Fig.~\ref{model}a.  Watts and Strogatz examined numerically the
average distance between pairs of vertices on small-world graphs and found
that only a small density of shortcuts is needed to produce distances
comparable to those seen in true random graphs.  At the same time, the
model shows the clustering which is characteristic of real social networks.

In this paper we derive an analytic solution for the distribution of path
lengths in the small-world model.  To do this we make use of a mean-field
approximation in which the distributions of quantities over the randomness
are represented by the average values of those quantities.  However, as we
will show, the mean-field theory turns out to be exact in the limit of
large system size for the small-world model.

The approach we use is first to solve the continuum version of the
Watts--Strogatz model shown in Fig.~\ref{model}b.  In this version of the
model the underlying one-dimensional lattice is treated as a continuum, and
one can measure the distance between any two points on this continuum,
rather than between only a discrete set of lattice sites.  Shortcuts are
assumed to have length zero.
%Fig.~\ref{model}b shows an example of a continuum small-world graph.
%Although it was never explicitly defined, the continuum
%small-world model was studied previously by Newman and Watts\cite{NW99a},
%who gave an exact equation for the distributions of path lengths on
%infinite discrete small-world graphs, but then solved this equation by
%making integral approximations to the sums they contained (see also
%Ref.~\cite{Moukarzel99}).  An entirely equivalent procedure would be to
%start with the continuum model and directly derive the integral forms.
Once we have a solution for the continuum model, we then note that if the
density of shortcuts is low the discrete and continuum models are
equivalent, and hence our solution is also a solution of the discrete
small-world model.

Consider then a ``neighborhood'' of radius $r$ centered around a randomly
chosen point on a small-world network of $L$ sites, where by neighborhood
we mean the set of points which can be reached by following paths of length
$r$ or less on the graph.  Let $m(r)$ be the number of sites on the graph
which do {\em not\/} belong to this neighborhood, averaged over many
realizations of the randomness in the graph, and $n(r)$ be the average
number of ``gaps'' around the lattice amongst which those sites are
divided---see Fig.~\ref{model}b.  Equivalently, $n(r)$ can be viewed as the
number of ``clusters'' of occupied sites.  In the continuum model both $m$
and $n$ are real numbers.  We will also find it convenient to use the
rescaled variables
\begin{equation}
\mu(r) = {m(r)\over L},\qquad \nu(r) = {n(r)\over L}.
\end{equation}

In the continuum limit the quantities $m(r)$ and $n(r)$ satisfy
differential equations as follows.  The rate at which the number of empty
sites on the lattice decreases with increasing $r$ is equal to the number
$2n$ of growing edges of clusters on the lattice times the range $k$ of
connections on the lattice.  Thus
\begin{equation}
{\d m\over\d r} = -2kn,
\end{equation}
or
\begin{equation}
{\d\mu\over\d r} = -2k\nu.
\label{diffeq1}
\end{equation}
This equation is exact for all values of $L$ and $\phi$.

The rate at which the number of gaps $n$ changes has two contributions.
First, the number of gaps increases as a result of the shortcuts on the
graph.  If $\xi=1/(k\phi)$ is the characteristic length defined in
Ref.~\onlinecite{NW99a} such that $L/\xi$ is the average number of
shortcuts in the graph, then the density of the ends of shortcuts on the
lattice is $2/\xi$.  This means that as $r$ increases, new shortcuts are
encountered at a rate $4kn/\xi$.  For each shortcut encountered, a new
cluster will be started at a random position on the lattice, provided that
the other end of the shortcut in question falls in one of the gaps around
the ring.  The probability of this happening is $m/L$.  Thus the rate at
which clusters (or gaps) are created is $4kmn/\xi L$.

The number of gaps decreases when the edges of a gap meet one another.
This will happen in the interval from $r$ to $r+\delta r$ if the size of
one of the gaps is less than $2k\,\delta r$.  If we consider all possible
ways of distributing the $m$ empty sites over $n$ gaps, we can see that the
probability distribution of the sizes of the gaps is the same as the
distribution of the smallest of $n-1$ uniformly distributed random numbers
$x$ between $0$ and $m$, which is
\begin{equation}
p(x) = {n-1\over m} \Biggl[1-\frac{x}{m}\Biggr]^{n-2}.
\end{equation}
Thus the probability of one particular gap being smaller than $2k\,\delta
r$ is $1 - (1-2k\,\delta r/m)^{n-1}$, which tends to $2k(n-1)\,\delta r/m$
in the limit of small $\delta r$.  The probability that any one of them is
smaller than $2k\,\delta r$ is $n$ times this.  Thus our final equation for
the rate of change of $n$ is
\begin{equation}
{\d n\over\d r} = {4kmn\over\xi L} - {2kn(n-1)\over m},
\end{equation}
or
\begin{equation}
{\d\nu\over\d r} = {4k\mu\nu\over\xi} - {2k\nu(\nu-1/L)\over\mu}.
\label{diffeq2}
\end{equation}
This equation is only exact when the average values $\mu(r)$ and $\nu(r)$
accurately represent the actual values of these quantities in the
particular realization of the model we are looking at, i.e.,~when the
distribution of values is sharply peaked.  This will be the case when the
number of shortcuts on the lattice is either much less than
one---$L\ll\xi$---or much greater than one---$L\gg\xi$---and therefore also
in the limit of large system size.  We have confirmed this using numerical
simulations, which show the distributions of $\mu$ and $\nu$ to be sharply
peaked in these limits but broad elsewhere.

Between them, Eqs.~\eref{diffeq1} and~\eref{diffeq2} are the fundamental
equations which lead to our solution for the small-world model.  As
demonstrated in the appendix, these equations can also be derived by
writing down difference equations for the variables $m$ and $n$ in the
discrete small-world model and then expanding in powers of the shortcut
density $\phi=1/\xi$ and keeping only the leading order terms.

We solve Eqs.~\eref{diffeq1} and~\eref{diffeq2} as follows.  First, we take
their ratio, which eliminates the variable $r$ and gives us a single
differential equation directly relating $\mu$ and $\nu$ thus:
\begin{equation}
{\d\nu\over\d\mu} = -{2\mu\over\xi} + {\nu-1/L\over\mu}.
\label{diffeq3}
\end{equation}
The general solution of this equation is
\begin{equation}
\nu = -{2\mu^2\over\xi} + {1\over L} + C\mu,
\end{equation}
where $C$ is an integration constant.  The constant can be fixed using the
boundary conditions $\mu(0) = 1$, $\nu(0) = 1/L$, which imply that
$C=2/\xi$ and hence
\begin{equation}
\nu = {2\over\xi} (\mu - \mu^2) + {1\over L}.
\end{equation}
Substituting this solution back into Eq.~\eref{diffeq1}, we get
\begin{equation}
{\d\mu\over\d r} = {4k\over\xi} (\mu^2 - \mu) + {2k\over L}.
\label{logistic}
\end{equation}
If we neglect the constant term in this equation, we arrive at the normal
logistic growth equation\cite{Strogatz94}, which will give an accurate
solution for $\mu$ in the regime where the lattice is neither very full nor
very empty.  If we keep all the terms, the general solution given the
boundary conditions is
\begin{eqnarray}
r &=& {\xi\over4k}
    \int_1^{\mu} {\d z\over z^2-z-\xi/2L}\nonumber\\
  &=& {\xi\over2k\sqrt{1+2\xi/L}}\,\Biggl[ \tanh^{-1}
    {1\over\sqrt{1+2\xi/L}} -
    \tanh^{-1} {2\mu-1\over\sqrt{1+2\xi/L}} \Biggr].
\label{solution1}
\end{eqnarray}
Rearranging for $\mu$ this gives
\begin{equation}
\mu = \frac{1}{2} \Biggl[ 1 + \sqrt{1+2\xi/L} \tanh \Biggl(
  \tanh^{-1} {1\over\sqrt{1+2\xi/L}} -
  2\sqrt{1+2\xi/L}\,{kr\over\xi} \Biggr) \Biggr].
\label{solution2}
\end{equation}
This equation gives us $\mu$ in terms of $r$, $\xi$ and $L$ for the
continuum version of the small-world model.  In the case where the typical
lattice distance between the ends of shortcuts is much larger than
one---$\xi\gg1$---the continuum version becomes equivalent to the normal
discrete version of the model and so in this limit our solution is also a
solution of the discrete small-world model.  Combining this condition with
the conditions specified earlier, we see that our solution will be exact
when either $1\ll L\ll\xi$, or when $1\ll\xi\ll L$.  This latter regime is
precisely the regime in which the small-world model is physically
interesting: the regime of large system size and large number of sparsely
distributed shortcuts.  In the intermediate regime between the two
conditions given, the solution is still quite accurate, and gives a good
guide to the general behavior of the model, as we will shortly show.

We now derive some of the more important consequences of
Eq.~\eref{solution2}.  First, we check that it reduces to the correct
expression in the case $L\to\infty$.  Making use of the identity
\begin{equation}
\tanh (x_1+x_2) = {\tanh x_1 + \tanh x_2\over1 + \tanh x_1 \tanh x_2},
\end{equation}
and we find that to first order in $\xi/L$
\begin{equation}
\mu = 1 + {\xi\over2L} [1 - \e^{4kr/\xi}],
\end{equation}
which agrees with the direct derivation for the $L=\infty$ case in
Ref.~\onlinecite{NW99a}.

Next, we note that once we have the fraction $\mu$ of sites not belonging
to a neighborhood of radius $r$, we can also calculate the number $A(r)\>\d
r$ in an interval from $r$ to $r+\d r$---the ``surface area'' of the
neighborhood---from
\begin{equation}
A = -L{\d\mu\over\d r} = 2 + {4L\over\xi} (\mu - \mu^2).
\label{area}
\end{equation}
Thus, once we have $\mu$ we can easily calculate $A$.

We can also derive an expression for the average vertex--vertex separation
$\ell$ on the graph, a quantity which has been studied by many
authors\cite{WS98,NW99a,Moukarzel99,BA99,NW99b,MMP99,BW99}.  We write
\begin{equation}
\ell = {1\over L} \int_0^\infty rA(r)\>\d r = \int_0^1 r\>\d\mu,
\label{ell1}
\end{equation}
where we have made use of Eq.~\eref{area}.  Even before performing the
integral, we can see that this implies certain behavior on the part of
$\ell$.  Eq.~\eref{solution1} shows that $kr/\xi$ is a function only of
$\mu$ and of the ratio of $L$ and $\xi$.  In other words $r$ has the form
\begin{equation}
r = {\xi\over k}\,h(\mu,L/\xi),
\label{rscale}
\end{equation}
where $h(x,y)$ is a universal scaling function with no dependence on the
parameters of the model other than through its arguments.  Substituting
this form into Eq.~\eref{ell1} and performing the integral over $\mu$, we
get
\begin{equation}
\ell = {\xi\over k}\,g(L/\xi),
\end{equation}
where $g(x)$ is another universal scaling function.  Except for the leading
factor of $1/k$, this scaling form is identical to the one suggested
previously by Barth\'el\'emy and Amaral\cite{BA99}.  Making the
substitution $g(x) = xf(x)$, we can also write it in the form
\begin{equation}
\ell = {L\over k} f(L/\xi) = {L\over k} f(Lk\phi),
\end{equation}
a form which was proposed by Newman and Watts on the basis of
renormalization group arguments, and which has been confirmed by extensive
numerical simulation\cite{NW99b,MMP99}.

\begin{figure}[t]
\begin{center}
\psfig{figure=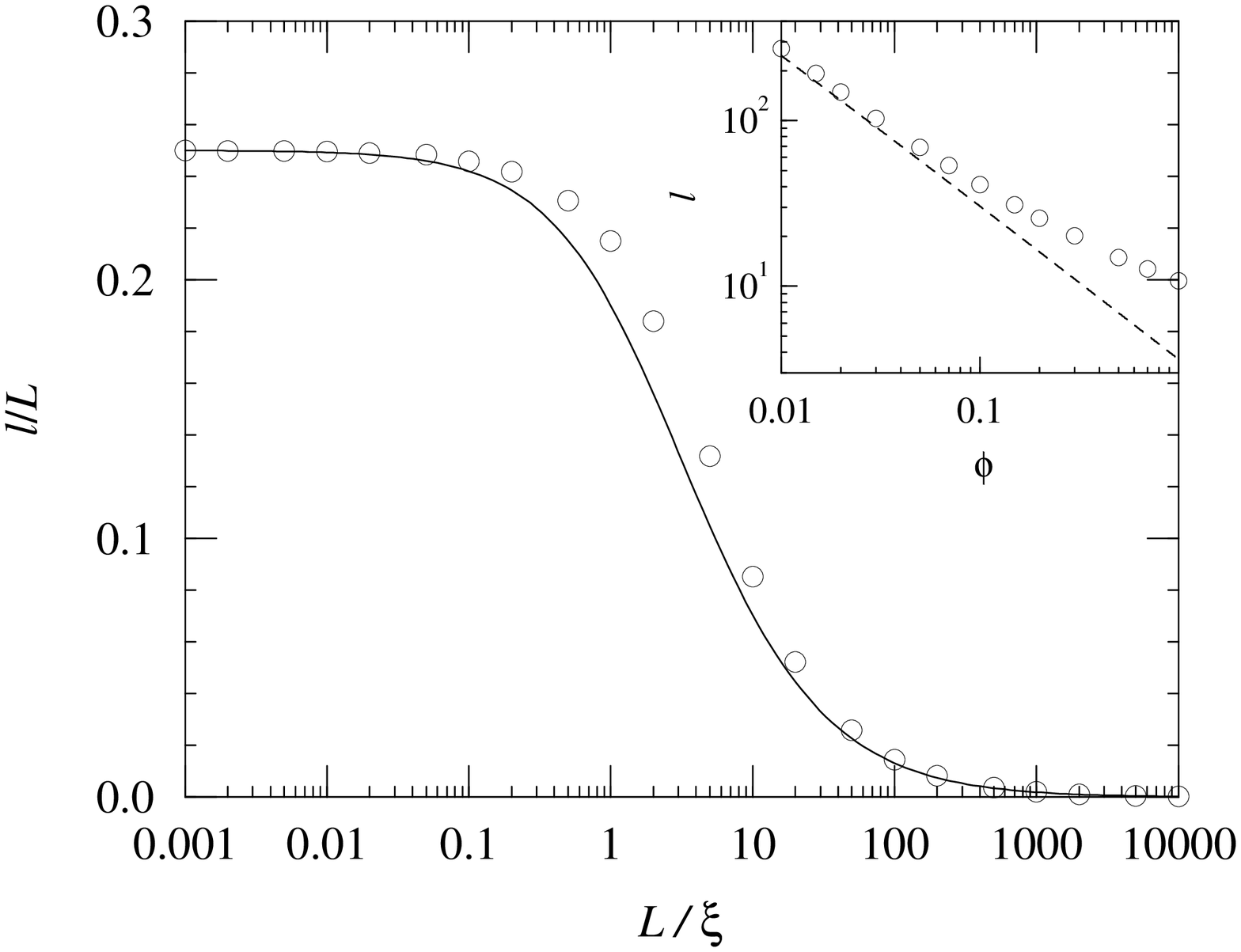,width=5in}
\end{center}
\capt{The average path length as a fraction of system size on a $k=1$
  small-world graph, plotted against the average number $L/\xi$ of
  shortcuts.  The circles are numerical measurements for the discrete model
  and the solid line is the analytic solution for the continuum model,
  Eq.~\eref{fscale}.  The error bars on the numerical measurements are
  smaller than the points.  Inset: the average path length on small-world
  graphs with $L=10^6$, for values of $\phi$ from $0.01$ up to $1$
  (circles) and the analytic solution, Eq.~\eref{ell2} (dotted line).}
\label{scaling}
\end{figure}

The complete solution for $\ell$ is obtained by
substituting~\eref{solution1} into~\eref{ell1} and performing the integral,
which gives
\begin{equation}
\ell = {\xi\over2k\sqrt{1+2\xi/L}} \tanh^{-1} {1\over\sqrt{1+2\xi/L}}.
\label{ell2}
\end{equation}
The scaling function $f(x)$ is then given by
\begin{equation}
f(x) = {1\over2\sqrt{x^2+2x}} \tanh^{-1} \sqrt{x\over x+2}.
\label{fscale}
\end{equation}
In Fig.~\ref{scaling} we show this form for the scaling function along with
numerical data from direct measurements of the average path length on $k=1$
discrete small-world graphs of size up to $L=10^7$ sites.  As the figure
shows, the two are in good agreement for large and small values of the
independent variable $x$ but, as expected, there is some disagreement in
the region around $x=1$ where $\xi$ and $L$ are of the same order of
magnitude.

The asymptotic forms of Eq.~\eref{fscale} are
\begin{equation}
f(x) \sim \left\lbrace \begin{array}{ll}
            \mbox{$\frac14$}& \qquad\mbox{for $x\ll1$}\\
            (\log 2x)/4x    & \qquad\mbox{for $x\gg1$,}
          \end{array} \right.
\label{limits}
\end{equation}
where we have made use of the identity
\begin{equation}
\tanh^{-1} x = \half\log{1+x\over1-x}.
\end{equation}
These forms are in agreement with previous conjectures\cite{BA99,NW99b},
which suggested that $f(x)$ should have the value $\frac14$ for small $x$
and should go as $(\log x)/x$ for large $x$.  As we see, the leading
numerical factor in the latter case is $\frac14$; this figure is exact,
since Eq.~\eref{solution2} is exact for large $L/\xi$.

In passing, we note that there is a simple physical interpretation of the
scaling function $f(x)$: apart from the leading factor of $\frac14$, it is
the fraction by which the average path length on a small-world graph is
reduced if the graph has $x$ shortcuts.  For example, Eq.~\eref{fscale}
indicates that it takes $x=3.5$ shortcuts on average to reduce the mean
path length by a half, and $44$ shortcuts to reduce it by a factor of
ten.  Thus only a small number of shortcuts are needed to reduce path
lengths quite considerably.  The same conclusion has been reached by Watts
and Strogatz\cite{WS98} on the basis of numerical data.

In the inset of Fig.~\ref{scaling} we show how our solution fails when the
shortcut density becomes too high.  The figure shows numerical results for
$\ell$ for a variety of values of the shortcut density $\phi=1/\xi$ from
$0.01$ up to $1$, for systems of one million sites (circles).  The dotted
line is Eq.~\eref{ell2}.  As the figure shows, the analytic solution is a
reasonable guide to the behavior of $\ell$ up to quite large values of
$\phi$ but, as expected, fails when $\phi$ gets close to~$1$.

To conclude, we have given a mean-field-like analytic solution for the
distribution of path lengths in the continuum version of the
Watts--Strogatz small-world model.  This solution is exact in the limit of
large system sizes for a given density of shortcuts, or in the limit of low
shortcut density for given system size.  In the case where the shortcut
density is low but the total number of shortcuts on the lattice is large
(because the lattice itself is also large) our solution is also an exact
solution of the normal discrete small-world model.  We have also derived an
expression for the average path length in the model and from this extracted
the scaling forms which this path length obeys.  We have checked our
results against numerical simulation of the discrete small-world model and
find good agreement in the regions where our solution is expected to be
exact.  In other regions the solution is a good guide to the general
behavior of the model but shows some deviation from the numerical results.

\newpage\appendix
\section*{The continuum model as the\\
small-$\phi$ limit of the discrete model}
In this appendix, we rederive Equations~\eref{diffeq1} and~\eref{diffeq2}
from the behavior of the discrete version of the small-world model to
leading order in the shortcut density $\phi$.

Consider a neighborhood of sites which are within distance $r$ of a given
starting site in the discrete model.  By analogy with the continuous case,
let $m$ be the number of sites on the lattice which are {\em not\/} in this
neighborhood and $n$ be the number of ``gaps'' between clusters of occupied
sites around the ring.  In fact, in the spirit of our mean-field
approximation, $m$ and $n$ should be thought of as the average of these
quantities over all possible realizations of lattice.  This means that they
may have non-integer values.  Here we treat them as integers for
combinatorial purposes, but our formulas are easily extended to non-integer
values by replacing the factorials by $\Gamma$-functions.

When we increase $r$ by one, the value of $m$ decreases for two reasons:
first because the gaps between clusters shrink and second because of new
sites which are reached by traveling down shortcuts encountered on the
previous step.  We write
\begin{equation}
m' = m - \Delta m
\label{deltam}
\end{equation}
where
\begin{equation}
\Delta m = \Delta m_g + \Delta m_s,
\label{delm1}
\end{equation}
with the two terms representing the shrinking of gaps and the shortcut
contribution respectively.

To calculate $\Delta m_g$, we note that the probability of any particular
gap having size~$j$ is
\begin{equation}
p_j = {{m-j-1\choose n-2}\over{m-1\choose n-1}},
\end{equation}
and the average number of such gaps is $n$ times this.  When we increase
$r$ by one, gaps of size $2k$ or larger shrink by $2k$, while gaps smaller
than $2k$ vanish altogether.  Thus
\begin{equation}
\Delta m_g = n \left[ \sum_{j=1}^{2k} j p_j 
           + 2k (1 - \sum_{j=1}^{2k} p_j) \right] 
           = m - \frac{(m-2k)! \,(m-n)!}{(m-1)! \,(m-n-2k)!}.
\label{delm2}
\end{equation}

To calculate the contribution $\Delta m_s$ from the number of shortcuts, we
note that the probability of encountering the end of a shortcut at any
given site is $2/\xi=2k\phi$, just as in the continuum case, so the number
of new shortcuts encountered when we increase $r$ by one is $2k\phi\,\Delta
m$.  $\Delta m_s$ in fact depends on the number of shortcuts encountered on
the {\em previous\/} increase in $r$, so we need to write $2k\phi\,\Delta
m^{(r-1)}$.  Only those shortcuts which land us at one of the $m-\Delta
m_g$ unoccupied sites contributes to $\Delta m_s$, so
\begin{equation}
\Delta m_s = 2k\phi\,\Delta m^{(r-1)} [m - \Delta m_g^{(r)}].
\label{seqn}
\end{equation}

Substituting Eqs.~\eref{delm2} and~\eref{seqn} into Eq.~\eref{delm1} we
get our complete expression for $\Delta m$.  Now we note that, except when
the lattice is very nearly full, the number of unfilled sites $m$ is of
order $L$.  The number of clusters of filled sites can be no greater than
the number of shortcuts on the lattice plus one for the initial cluster
around the starting point.  Thus $n \le \phi L + 1$, and the ratio
$(n-1)/m$ is a quantity of order $\phi$.  Expanding in powers of this
quantity and assuming the number of sites $m$ to be much greater than $2k$
then gives
\begin{equation}
\Delta m = 2 k n,
\label{delmfinal}
\end{equation}
plus terms of order $\phi$.  Physically, the reason for the simplicity of
this expression is that, to first order in $(n-1)/m$, most gaps have size
$2k$ or larger, and the contribution from new shortcuts to can be
neglected, since most sites are connected only to their local neighbors.

The change in the value of $n$ has also two contributions.  First, the
number of gaps increases when a shortcut creates a new cluster, and divides
a gap into two new ones.  As we have already shown this happens $\Delta
m_s$ times on average when we increase $r$ by one, where $\Delta m_s$ is
given by Eq.~\eref{seqn}.  Second, gaps disappear when their edges meet.
When we increase $r$ by one, a gap will close if its size is $2k$ or less.
Thus the change in $n$ is
\begin{equation}
n' = n + \Delta n,
\end{equation}
where
\begin{equation}
\Delta n = \Delta m_s - n \sum_{j=1}^{2k} p_j 
 = \Delta m_s - n \, \frac{(m-n)! \,(m-2k-1)!}{(m-1)! \,(m-n-2k)!} .
\label{deltan}
\end{equation}
where $\Delta m_s$ is given by Eq.~\eref{seqn}.  Expanding to lowest order
in $(n-1)/m$, taking $m\gg2k$ again, and combining the result with
Eq.~\eref{seqn} gives
\begin{equation}
\Delta n =  {4k^2 \phi m n\over L} - {2kn(n-1)\over m}.
\label{delnfinal}
\end{equation}

Changing Eqs.~\eref{delmfinal} and \eref{delnfinal} from difference
equations to differential ones and dividing by $L$ to transform from
$m$ and $n$ to $\mu$ and $\nu$ gives
\begin{eqnarray}
{d\mu \over dr} & = & - 2 k \nu,\\
{d\nu \over dr} & = & 4 k^2 \phi\mu\nu + \frac{2 k \nu(\nu-1/L)}{\mu}.
\end{eqnarray}
Recalling that $\xi = 1/(k\phi)$, we can see that these equations are
identical to Equations~\eref{diffeq1} and~\eref{diffeq2}.

\end{document}